\documentclass[3p,times,twocolumn]{elsarticle}
\usepackage{graphicx}
\usepackage{color}
\newcommand{\mib}{$\mu_{B}$ }
\newcommand{\mibn}{$\mu_{B}$}
\newcommand{\ptopi}{$p/\rm \pi$ }

\newcommand{\ptopip}{$p/\rm \pi^{+}$ }
\newcommand{\ptopipn}{$p/\rm \pi^{+}$}
\newcommand{\aptopi}{$\bar{p}/\rm \pi^{-}$ }
\newcommand{\aptopin}{$\bar{p}/\rm \pi^{-}$}
\newcommand{\rootsnn}[1]{$\sqrt{s_{NN}} = #1$~GeV}
\newcommand{\rootsnp}[1]{$\sqrt{s_{NN}} = #1$ }
\newcommand{\roots}[1]{$\sqrt{s} = #1$~GeV}

\newcommand{\pt}{$p_{T}$ }
\newcommand{\ptn}{$p_{T}$}
\newcommand{\pteq}[1]{$p_{T} = #1$~GeV/$c$}
\newcommand{\ptgt}[1]{$p_{T} > #1$~GeV/$c$}
\newcommand{\ptapp}[1]{$p_{T} \approx #1$~GeV/$c$}

\begin{document}
\begin{frontmatter}
\title{Rapidity dependence of the proton-to-pion ratio in Au+Au
  and p+p collisions at \rootsnp{62.4} and 200~GeV\tnoteref{t1}}
\tnotetext[t1]{The BRAHMS Collaboration}

\author[oslo]{I.~G.~Arsene\fnref{fn5}}
\author[nbi]{I.~G.~Bearden}
\author[bnl]{D.~Beavis}
\author[kansas]{S.~Bekele\fnref{fn10}}
\author[bucharest]{C.~Besliu}
\author[newyork]{B.~Budick}
\author[nbi]{H.~B{\o}ggild}
\author[bnl]{C.~Chasman}
\author[nbi]{C.~H.~Christensen}
\author[nbi]{P.~Christiansen\fnref{fn3}}
\author[nbi]{H.~H.~Dalsgaard}
\author[bnl]{R.~Debbe}
\author[nbi]{J.~J.~Gaardh{\o}je}
\author[texas]{K.~Hagel}
\author[bnl]{H.~Ito}
\author[bucharest]{A.~Jipa}
\author[kansas]{E.~B.~Johnson\fnref{fn4}}
\author[nbi]{C.~E.~J{\o}rgensen\fnref{fn9}}
\author[krakow]{R.~Karabowicz}
\author[krakow]{N.~Katry{\'n}ska}
\author[bnl,kansas]{E.~J.~Kim\fnref{fn8}}
\author[oslo]{T.~M.~Larsen}
\author[bnl]{J.~H.~Lee}
\author[oslo]{G.~L{\o}vh{\o}iden}
\author[krakow]{Z.~Majka}
\author[krakow]{A.~Marcinek}
\author[texas,kansas]{M.~J.~Murray}
\author[texas]{J.~Natowitz}
\author[nbi]{B.~S.~Nielsen}
\author[nbi]{C.~Nygaard}
\author[kansas]{D.~Pal}
\author[oslo]{A.~Oviller}
\author[krakow]{R.~P\l aneta}
\author[ires]{F.~Rami}
\author[nbi]{C.~Ristea}
\author[bucharest]{O.~Ristea}
\author[bergen]{D.~R{\"o}hrich}
\author[kansas]{S.~J.~Sanders}
\author[krakow]{P.~Staszel \fnref{fn1}}
\author[oslo]{T.~S.~Tveter}
\author[bnl]{F.~Videb{\ae}k \fnref{fn2}}
\author[texas]{R.~Wada}
\author[bergen]{H.~Yang\fnref{fn6}}
\author[bergen]{Z.~Yin\fnref{fn7}} 
\author[bucharest]{and I.~S.~Zgura}

\fntext[fn1]{Corresponding author. Email: ufstasze@if.uj.edu.pl}
\fntext[fn2]{Spokesperson. Email: videbaek@bnl.gov}
\fntext[fn5] {Current Address: ExtreMe Matter Institute EMMI, GSI Helmholtzzentrum f\"{u}r
Schwerionenforschung GmBH, Darmstadt, Germany}
\fntext[fn10]{Current Address: Dept of Physics, Tennessee Tech University,Cookeville, Tennessee 38505}
\fntext[fn3]{Current Address: Div. of Experimental High-Energy
Physics, Lund University, Lund, Sweden}
\fntext[fn9]{Ris\o \  National Laboratory, Denmark}
\fntext[fn8] {Current Address: Division of Science Education,
  Chonbuk National University, Jeonju, 561-756, Korea}
\fntext[fn4] {Current Address: Radiation Monitoring Devices, Cambridge, Massachusetts} 
\fntext[fn6] {Current Address: Physics Institute, University of Heidelberg, Heidelberg, Germany}
\fntext[fn7] {Current Address: Institute of Particle Physics, Huazhong Normal
   University, Wuhan, China}

\address[bnl]{Brookhaven National Laboratory, Upton, New York 11973}
\address[ires]{Institut Pluridisciplinaire Hubert Curien CRNS-IN2P3 et
  Universit{\'e} Louis  Pasteur, Strasbourg, France}
\address[newyork]{New York University, New York 10003}
\address[nbi]{Niels Bohr Institute, Blegdamsvej 17, University of Copenhagen, Copenhagen 2100, Denmark}
\address[krakow]{Smoluchowski Inst. of Physics, Jagiellonian University, Krak\'ow, Poland}
\address[texas]{Texas A$\&$M University, College Station, Texas, 77843}
\address[bergen]{University of Bergen, Department of Physics and Technology, Bergen, Norway}
\address[bucharest]{University of Bucharest, Romania}
\address[kansas]{University of Kansas, Lawrence, Kansas 66045}
\address[oslo]{University of Oslo, Dep. of Physics, Blindern, 0316 Oslo, Norway}

\begin{abstract}

The proton-to-pion  
ratios measured in the BRAHMS experiment  
for Au+Au and
p+p collisions at $\sqrt{s_{NN}}$ = 62.4 and 200~GeV
are presented as a function of transverse momentum and collision centrality
at selected pseudorapidities in the range of 0 to 3.8. 
A strong pseudorapidity dependence of these ratios 
is observed.
We also compare the magnitude and \ptn-dependence of the \ptopipn~ ratios measured in 
Au+Au collisions at \rootsnn{200} and  $\eta \approx 2.2$ with the same ratio measured 
at \rootsnn{62.4} and $\eta = 0$. The great similarity  
found between these ratios
throughout the whole \pt range (up to $2.2$~GeV/$c$) is consistent with particle ratios in A+A collisions
being described with grand-canonical distributions
characterized by the baryo-chemical potential  \mibn.
At the collision energy of 62.4~GeV, we have observed a unique point
in pseudorapidity, $\eta = 3.2$, where the \ptopip ratio is independent of
the collision system size in a wide \ptn-range of $0.3 \le p_{T} \le 1.8$~GeV/$c$.

\end{abstract}
\end{frontmatter}

\section{Introduction}

The ongoing dialogue between theory and experiment is delivering 
an increasingly clear
picture of the QCD phase diagram
in its partonic and hadronic phases. In particular, recent mid-rapidity RHIC results are
considered as evidence for a smooth cross over from a dense and opaque partonic phase into
a hadronic gas at temperatures around 170 MeV and small values of baryo-chemical potential \mib \cite{stephanov}.   
Measurements of the relative abundances of hadronic species  constrain statistical
models used to describe  
the chemical freeze-out in nucleus-nucleus interactions
at different energies. Some of these models show a remarkable behavior 
at low baryo-chemical potential where the curves of temperature vs. baryo-chemical potential
representing chemical freeze-out tend to merge with the phase boundary  between partonic and 
hadronic media 
 \cite{braun_m}. In such a picture, hadrons are produced very close to freeze-out, and one could safely state
that the features of the partonic medium are transmitted
to the final bulk hadrons via hadronization processes.   
Indeed, such appears to be the case for the constituent quark scaling found in elliptic 
flow $v_{2}$ measurements \cite{elliptic_flow} as well as the enhancement of baryon-to-meson ratios
that scale with the collision size around mid-rapidity (low \mibn) \cite{ejkm, qm2008}. 
This increase in the baryon-to-meson ratios ($\approx 1$ for the so-called intermediate
\pt values ranging from 2 to 6~GeV/$c$), first reported by the PHENIX Collaboration \cite{PHENIX_Adcox},
deviates remarkably from calculations 
that include parton fragmentation in the vacuum. 
The \ptn-dependence of the baryon-to-meson ratio in the intermediate
\pt range appears to be related to a modification of hadronization in  
the partonic medium created by the collision. This effect could be  
caused by the different quark content of the baryons and mesons \cite{rj_fries}  
or because of their different masses through the effect of radial  
flow \cite{peitzmann}.
 
Both radial flow and in-medium quark coalescence are expected to 
enhance protons over pions at intermediate \ptn.
In particular, the PHENIX \aptopi data at mid-rapidity 
is well described by the  Greco, Ko, and Levai quark coalescence model
\cite{GKL} where the introduced
coalescence involves partons from the medium (thermal) and partons from mini-jets.
The Hwa and Yang quark recombination model \cite{hwa_67}  
describes well the BRAHMS and PHENIX \ptopip ratios around mid-rapidity.
These mid-rapidity results are consistent with hadronization in the intermediate \pt range 
being dominated by parton recombination with negligible final state interactions
between the produced hadrons.
In contrast, at forward rapidity (large $\mu_{B}$), a significant gap
between the temperature of the transition from the partonic to the
hadronic phase, $T_{c}$, and the temperature of chemical freeze-out is
predicted by QCD lattice calculation \cite{LQCD_vs_freeze-out}. 
In that environment,  
hadronic re-scattering may play an important role, making statistical models more
suitable to 
describe particle abundances at high rapidity \cite{hirano, broniowski_flor}. 

We present in this Letter the \ptopi ratios for both charges extracted from Au+Au collisions at 
different centralities as well as p+p at $\sqrt{s_{NN}}$ = 62.4 and 200~GeV. The ratios are presented 
as function of pseudorapidity and transverse momentum \ptn. The system size dependence of the ratios
is investigated and compared to the smallest system; p+p collisions. The comparison of the \ptopip
measured at two different energies and pseudorapidities is used to establish a connection 
between particle abundances in these systems and their possible description by a grand-canonical
distribution. We also present an interesting feature in the pseudorapidity dependence of the \ptopip ratio
at $\sqrt{s_{NN}}$ = 62.4~GeV. There  the ratio has the same magnitude and \ptn~dependence at $\eta = 3.2$
when measured at all Au+Au centralities as well as p+p collisions. We compare the ratios extracted 
from the most central Au+Au events to a hydrodynamical model and to a partonic cascade model that 
includes hadronic re-scattering in the final state.
  
\section{Experimental setup}  

The BRAHMS detector setup \cite{brahms_det} consists of two movable, small acceptance
spectrometers: the Mid-Rapidity Spectrometer (MRS), which operates in the
polar angle interval from 
$90^{\circ} \ge \theta \ge 30^{\circ}$
(corresponding to
a pseudorapidity interval of $0 \leq \eta \leq 1.3$) and the Forward
Spectrometer (FS), which operates in the
 range from  
$15^{\circ} \ge \theta \ge 2.3^{\circ}$ 
($2\leq \eta \leq 4$). 
The BRAHMS setup also includes 
detectors used to determine global features of the collision such as the overall charged particle multiplicity,
the collision vertex and the centrality of the collision.
The MRS is composed of a single dipole magnet  
placed between two  Time Projection Chambers (TPC),
which provide a momentum measurement. 
Particle identification (PID) is based on Time-of-Flight (TOF) measurements \cite{brahms_det}.
The FS has two TPCs, 
which are capable of track recognition in a high multiplicity environment close to the interaction region, 
and, at the far end of the spectrometer, three Drift Chambers. 
In the aggregate, the FS can detect particle track segments with
high momentum resolution ($\delta p / p = 0.0008 p$ at the highest
field setting) using three dipole magnets.
Particle identification in the FS is provided by TOF measurements in two separate
hodoscopes for low  and medium  
particle momenta, respectively.
High momentum particles are identified using
a Ring Imaging Cherenkov detector (RICH) \cite{rich_nim}.

\section{The analysis}

The analysis reported in this Letter consists of the comparison of proton and 
pion yields as a function of \pt 
for several pseudorapidity intervals.
The analysis is carried out in $\eta$ versus \pt
  space, since for any given angle and field configuration 
the acceptance is the same for pions and protons.
For a given $\eta$-\pt bin the \ptopi ratios are calculated on a
setting by setting basis. 
In order to avoid mixing different PID techniques,
which can lead to different systematic uncertainties,  
the ratios are calculated separately for PID using TOF and the RICH detectors.
In this way, all factors such as acceptance corrections, 
tracking efficiencies, trigger normalization and bias related to the centrality cut
cancel out in the ratio.  
The remaining species dependent corrections are: 
\begin{enumerate}[(i)]
\item 
decays in flight and interactions with the beam pipe and other material, and
\label{enu1}
\item the PID efficiency correction. \label{enu2}
\end{enumerate} 

The corrections for (\ref{enu1}) are determined from 
the single particle response of pions and protons in
a realistic GEANT \cite{GEANT} model description of the BRAHMS experimental setup.
The magnitude of this correction on the particle ratios depends on the 
particle momenta and the spectrometer positions,
but does not exceed $6\%$. 
We estimate that the overall systematic uncertainty related to
this correction  is at the level of $2\%$.
 
The TOF PID is done separately for small momentum bins by fitting a
multi-Gaussian function to the experimental squared mass $m^{2}$
distribution and applying a $\pm 3 \sigma$ cut to select a given
particle type. For measurements done with the FS spectrometer in the 
momentum range where pions overlap with kaons
(usually above $3.5$~GeV/$c$), the RICH detector can be used in veto mode  
to select kaons  with momentum smaller than  the kaon Cherenkov threshold,
which is about $9$~GeV/$c$. 
  This procedure leads to a relatively clean sample of pions with some
contamination by kaons having spurious rings associated in the RICH counter.
Together with the kaon - proton overlap at larger momenta, this contamination
effect is a source of systematic errors which have been estimated 
to be in the order of a few percent for all \pt values, except for low \ptn in 
the \aptopi ratios, where these uncertainties can reach high values ($ \approx 15\%$).
At mid-rapidity
the systematic uncertainty reaches a value of $5\%$ at \ptgt{2.5} due
to the limited kaon to pion separation at large momenta. 

The RICH PID is also based on the particle separation in the
$m^{2}$ versus momentum space. The RICH provides direct proton
identification above the proton threshold momentum which is about $15$~GeV/$c$.
 However, an additional proton identification scheme is possible 
below this value but above the kaon
threshold.  
In this momentum range a proton is associated with tracks having 
momenta above the kaon threshold, but no RICH signal (veto mode).
The veto proton yields are corrected for pion and
kaon contamination due to a small RICH inefficiency.
The RICH inefficiency was determined by studying yields of tracks
identified in the TOF detector as pions but having no associated ring 
in the RICH. This study was done for a low momentum range
with a good kaon/pion separation in TOF. 
It is found that the pion efficiency grows rapidly from the pion 
threshold ($\approx 2.3$~GeV/$c$) and reaches
a constant value of about $97 \%$ above 4~GeV/$c$.
The RICH inefficiency found for pions is then applied to kaons (and
the other species) assuming that Cherenkov
radiation depends only on the $\gamma$ factor of the particle. 
A more detailed description of the RICH inefficiency analysis is given
in \cite{rich_nim} and a forthcoming publication. 
There are two sources of systematic uncertainties associated with the RICH
PID, namely, the uncertainty on the RICH inefficiency, 
estimated to be at the level of $10\%$, and the overlap in
$m^{2}$ between pions and kaons having momenta above about 30~GeV/$c$. 

No corrections were applied for weak decays. However
we apply cuts on track and event vertex matching to 
limit the effects of feed down on particle
yields.
The ranges of the vertex cuts are determined by the measured spatial
resolution of the particle track projection to the event vertex,
such that 97$\%$ of primary tracks are included.
Using AMPT \cite{AMPT:z-lin} model calculations we found that 
these vertex cuts lead to remnant contamination of proton and pion 
yields mainly 
due to $\Lambda$ hyperons
 and $K^{0}_{s}$  meson decays. 
At mid-rapidity we estimated
that this contamination leads to a 10$\%$(7$\%$) enhancement of
\ptopipn(\aptopin) at \pt = 3.0 GeV/$c$. This enhancement increases
toward low transverse momenta and reaches
20$\%$(14$\%$) for \ptopipn(\aptopin) at  \pt = 1.5 GeV/$c$.
The level of contamination decreases gradually towards 
forward rapidities and at
$\eta \approx 3$ the enhancement reaches values of 6$\%$(5$\%$) 
for \ptopipn(\aptopin) at \pt = 3.0 GeV/$c$ 
and 10$\%$(7$\%$) for \ptopipn(\aptopin) at  \pt = 1.5 GeV/$c$.  
We also found from the model calculations that the extracted
  contaminations are proportional to the primary $\Lambda / p$ ratios
  ($\bar{\Lambda} / \bar{p}$ ratios for negative species). Thus if the
  AMPT model under predicts the
data by 20$\%$ then the respective corrections due to contaminations should be
increased by approximately 20$\%$.

\section{Results}

\begin{figure}[t]
\includegraphics[width=0.5\textwidth,totalheight=0.43\textheight]
                {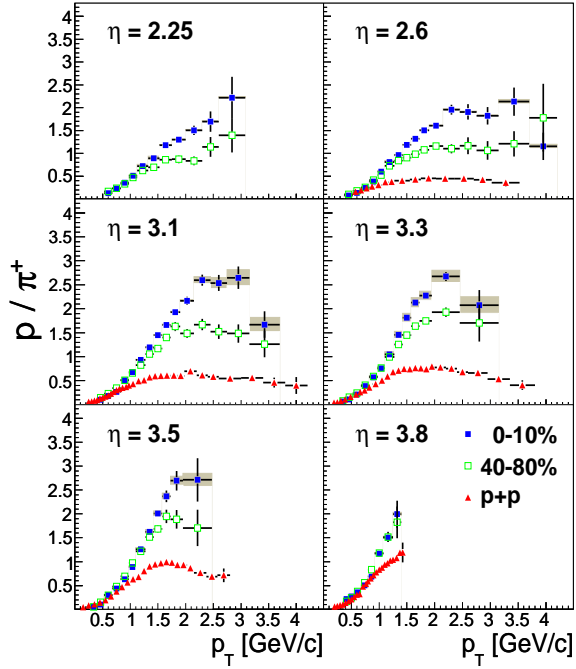}
\caption{Centrality dependent \ptopip ratios for Au+Au system  
colliding at \rootsnn{200} for central ($0-10\%$) and  
semi-peripheral ($40-80\%$) reactions in comparison with 
 p+p collision data at the same energy. The vertical bars represent the  
statistical errors and the shaded boxes (plotted only for central
  Au+Au) show the systematic uncertainties.}
\label{fig1}
\end{figure}
Figure \ref{fig1} shows \ptopip ratios 
measured in Au+Au collisions
at \rootsnn{200} 
in two centrality classes of events, namely, 
$0-10\%$ (solid dots) and $40-80\%$ (open squares).
The centrality selection is based on the charged-particle multiplicity
measured in the range $-2.2 < \eta <2.2$ as described in \cite{brahms_prl88}.
The shaded boxes, plotted for the most central events,
represent the systematic uncertainties discussed in the previous 
section. However, they do not include the uncertainties 
related to the weak decay contamination. 
The ratios extracted from p+p data at the same energy are plotted for comparison (solid triangles). 
The \pt coverage depends
on the pseudorapidity bins and for central Au+Au collisions 
extends up to \pteq{4} for $\eta = 2.6$. 
At low \pt ($ < 1.5$~GeV/$c$) the \ptopip ratios exhibit
a rising trend with a weak dependence on centrality.
A significant dependence on centrality begins above $1.5$~GeV/$c$.
The ratios appear to reach a
maximum value at \pt around 2.5~GeV/$c$ 
wherever the \pt coverage is sufficient to determine this limit.
The maxima of the ratios are greater for more central events and,
at $\eta = 3.1$, are equal to about 2.5 and 1.5
for the $0-10\%$ and $40-80\%$ centrality bins, respectively. 
The p+p ratios are consistent with Au+Au data at low \pt and begin to deviate 
significantly above \pteq{1}. At $\eta = 3.1$ a maximum value of the ratio
of 0.55 is reached in p+p collisions which is
a factor of 4.5 smaller than that observed for central Au+Au reactions. 
We have performed PYTHIA \cite{pythia} calculations  for nucleon-nucleon
interactions in each possible isospin state. 
The calculations of the \ptopi ratio vs. 
\pt show no significant dependence on the isospin of the initial NN system. Thus the 
difference between p+p and Au+Au is not an isospin effect.

\begin{figure}[t]
\includegraphics[width=0.5\textwidth,totalheight=0.43\textheight]
                {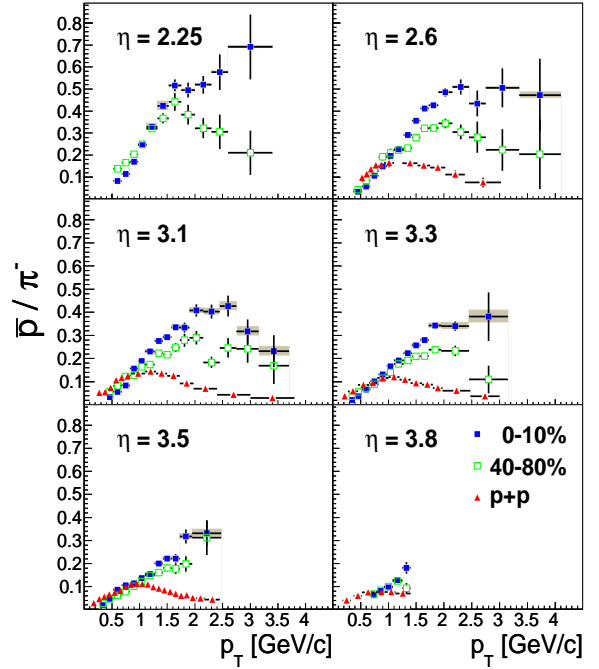}
\caption{ The  ratios of \aptopin measured in Au+Au and p+p 
collisions at \rootsnn{200}, are plotted for the same centrality samples used  in Fig. \ref{fig1}.}
\label{fig2}
\end{figure}

The values of the \aptopi ratios plotted in Fig. \ref{fig2} are significantly
lower than the \ptopip ratios (note the difference in the vertical
scale), however, the centrality dependence shows the same features
as those observed in the \ptopip ratios, namely,
that the ratios for different centrality classes
are consistent with each other up to \ptapp{1.2} and
a strong dependence on centrality appears at larger transverse momenta  
reaching a maximum at similar \pt as the positive particles.
Looking at the p+p  data alone, one notes the difference 
in shape between the  \ptopip and \aptopi ratios: a clear shift of the \aptopi
peaks towards lower \ptn, as well as a much broader \ptopip peaks. 
The large difference between the Au+Au and p+p
both in shape and overall 
magnitude may reflect significant medium effects in Au+Au collisions at
\rootsnn{200} in the pseudorapidity intervals covered.

\begin{figure}
\includegraphics[width=0.49\textwidth,totalheight=0.28\textheight]
                {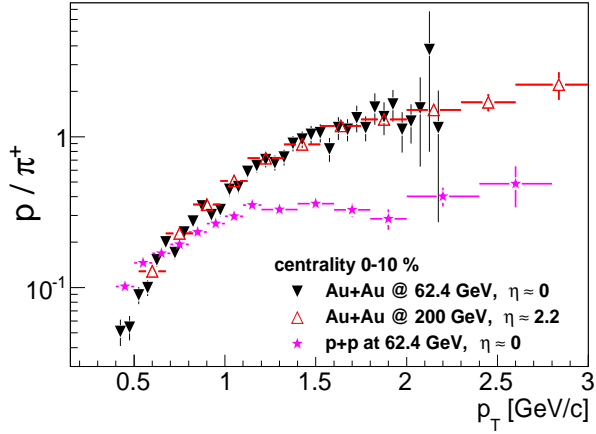}
\caption{\ptopip ratio in Au+Au collisions for $\eta =
0.0$ at \rootsnn{62.4} marked with open
triangles and the Au+Au reactions for  $\eta = 2.2$ at
\rootsnn{200} marked with the black triangles. 
Stars shown reference mid-rapidity data for p+p at \rootsnn{62.4}.}
\label{fig6}
\end{figure}

In Fig. \ref{fig6} we explore the possible connection between the measured \ptopi ratios in 
extended systems and the baryo-chemical potential $\mu_{B}$ used to characterize them as statistical systems. 
Such a connection is done by comparing two rapidity ranges that have  
the same $ \bar{p}/p$ ratio but different $\sqrt{s_{NN}}$. These are $\eta=0$ at 62 GeV  
and $\eta=2.2$ at 200 GeV.
The pseudorapidity intervals selected for this comparison correspond
to similar observed $\bar{p}/p$ ratios of approximately 0.45.
The very similar $\bar{p}/p$ ratios in these two systems has been attributed to their
having a common value of the baryo-chemical potential $\mu_{B}$ of $\sim 62$~MeV
~\cite{ratio200,ratio62}.
The similarity of proton-to-pion ratios for these selected
heavy ion collisions suggests that the baryon and meson production at the $p_{T}$ interval studied
(up to 2 GeV/$c$) is dominated by medium effects and is determined by
the bulk medium properties. 
The considerably lower values of the \ptopip
ratio measured in the p+p system at \roots{200}, shown with  
stars in Fig.~\ref{fig6}, can also be construed as strong 
indication that medium effects are the source of
the observed enhancement of the \ptopi as function 
of \pt in the nucleus-nucleus collisions.

\begin{figure}[th]
\includegraphics[width=0.48\textwidth,totalheight=0.57\textheight]
                {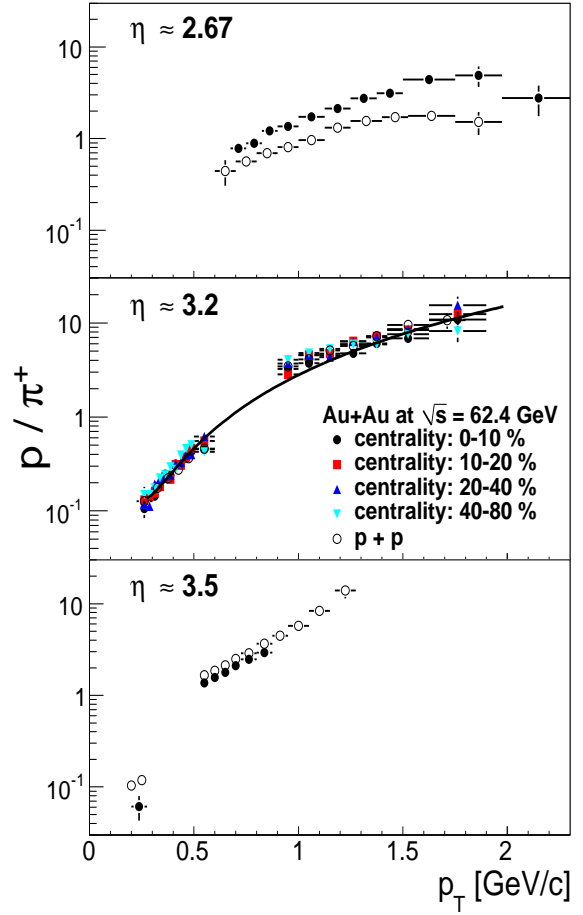}
\caption{\ptopip ratio from p+p and Au+Au collisions
at $\sqrt{s_{NN}}$ = 62.4~GeV for $\eta$ = 2.67 (top panel), $\eta$ =
3.2 (middle panel) and $\eta$ = 3.5 (bottom panel). The curve drawn in the middle panel is a 
polynomial
fit to the most central events, its purpose is to guide the eye and bridge the gap in \pt.}
\label{fig7}
\end{figure}

The three panels of Fig.~\ref{fig7} display the \ptopip ratio extracted from p+p 
and Au+Au collisions
at \rootsnn{62.4} measured at 3 different pseudorapidity bins;
$\eta$ = 2.67 (top panel), $\eta$ = 3.2 (middle panel) and $\eta$ = 3.5 (bottom panel). 
At $\eta=2.67$, \ptopip is greater in Au+Au than in p+p collisions while at  
$\eta=3.5$ the situation is reversed.
The middle panel of Fig.~\ref{fig7} shows the \ptopip ratios
from Au+Au and p+p collisions at \rootsnn{62.4}
at $\eta$ = 3.2. This "crossing point" in pseudorapidity shows a 
remarkably complete
overlap of the ratios as function of \pt not only for p+p and $0-10\%$ central Au+Au reactions, but 
also for other Au+Au centralities, namely $10-20\%$, $20-40\%$ and
$40-80\%$. 
Such a universal shape for the \ptopip ratios implies, 
that for each centrality at this pseudorapidity,
the nuclear modification factors for protons and pions are consistent
with each other at all measured values of \ptn.
The observed crossing of \ptopip ratios for different 
sizes of the colliding
systems is consistent with recent BRAHMS results on the centrality
dependence of net-proton rapidity distribution in Au+Au reactions at \rootsnn{200} 
\cite{fv_qm2009}.    
The data show an increased baryon transport towards mid-rapidity in central collisions. 
Such increased
stopping power dissipates the beam energy to form a
denser system where recombination mechanisms favor proton production
at intermediate values of \ptn. Both effects will produce higher values of 
the \ptopip ratio in larger systems.   
On the other hand, at very forward rapidities (around one unit below
the beam rapidity) a bigger fraction of the measured particles are protons 
and their numbers are even higher for lighter systems
due to reduced stopping power. 
\begin{figure*}[t]
\includegraphics[width=0.98\textwidth,totalheight=0.37\textheight]
                {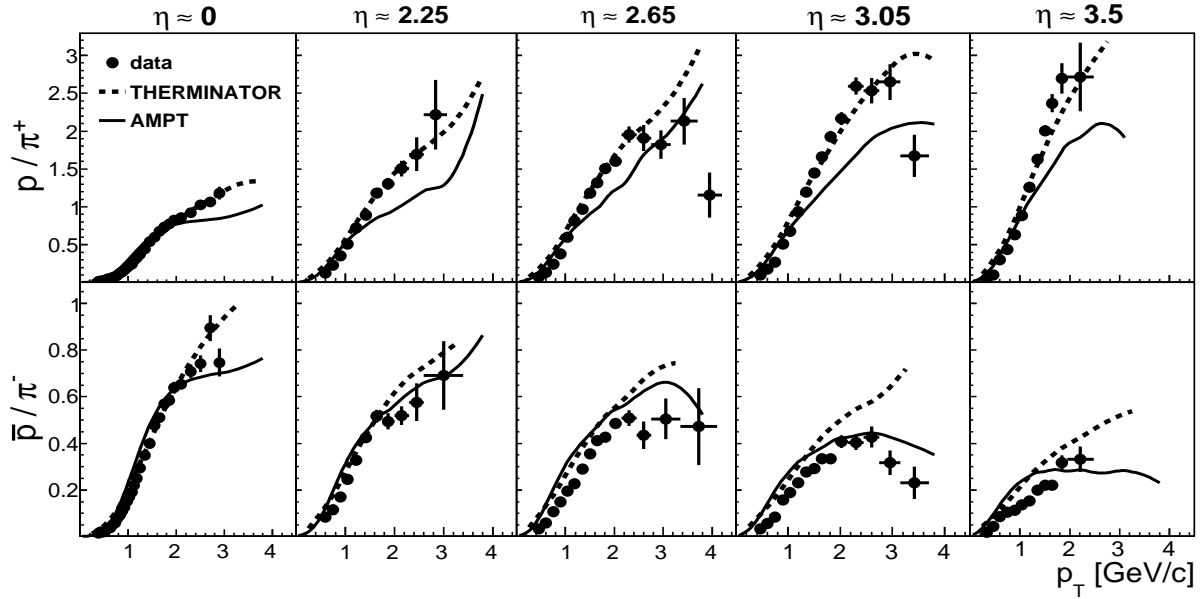}
\caption{Rapidity evolution of \ptopip and \aptopi for
$0-10\%$ central Au+Au reaction at \rootsnn{200} (solid points), and calculations
from the THERMINATOR (dashed line) and the AMPT models (solid line).}
\label{fig5} 
\end{figure*}
This observation is consistent with the reversed \ptopip
dependence on the system size seen in the bottom panel.  
A crossing of \ptopip yields ratio at the energy of 62.4~GeV is predicted by 
UrQMD \cite{URQMD}, HIJING \cite{hijing} and AMPT  model calculations.
However, these models predict the location of the crossing point in the interval
$2 < y < 2.5$, which is almost one unit
of rapidity lower than the observed value. These experimental results
then  provide a strong constraint on the theoretical
description of baryon number transport and associated energy
dissipation in relativistic nuclear reactions.
It is worth noting that at 200~GeV (see~Fig. \ref{fig1}), even  
for $\eta$ = 3.8, the \ptopip ratios
measured in Au+Au reactions are larger than those observed in p+p collisions, so
the possible crossing point at high energy is located at larger rapidity, beyond 
the experimental acceptance.

\section{Model comparisons}

To interpret these results, theoretical 
models of nucleus-nucleus collisions are confronted with the data.
The \ptopip ratio measured in central Au+Au collisions at
\rootsnn{200}, shown on the top row of
Fig. \ref{fig5}, has a strong growth with rapidity;  starting 
from a value of about 1.0 at $\eta \approx 0$ and \pt $\approx 3$~GeV/$c$ and reaching a value of
 2.5 at $\eta \approx 3.5$ and \pt $\approx 3$~GeV/$c$.
In contrast, the \aptopi ratio
(bottom row) decreases with increasing rapidity ( from $\approx 1$ at $\eta \approx 0$
to $0.4$ at $\eta \approx 3.5$). Note the different vertical scales for
positive and negative charged particles. 
Figure~\ref{fig5} compares our results
for $0-10\%$ central Au+Au reactions at \rootsnn{200}
to two model calculations based on: the THERMINATOR model \cite{broniowski} (black dashed lines) and the AMPT 
model  (solid lines). 
THERMINATOR is a 1+1-D hydrodynamic model that incorporates 
statistical particle production (including excited states), which in turn, 
has chemical potentials with parameterized rapidity dependence. 
Alternatively, the 
AMPT (A Multi-Phase Parton Transport) model  
includes mini-jet parton production, parton dynamics,
hadronization according to the LUND string fragmentation model,
and final state hadronic interactions in determining the final particle production. 
In these simulations we duplicate experimental conditions regarding the particle contamination
due to weak decays, the tracks were required to point to the 
event 
vertex under the same conditions as those applied to the experimental data.

The THERMINATOR model
tracks well the trends of the data up to \ptapp{2.5} for both positive and
negative particles. 
At $\eta = 0$ this model describes well the experimental
\ptopip and \aptopi ratios in the intermediate \pt range covered 
by the data ($1 <$ \pt $< 3$ GeV/$c$), however, it predicts the
\aptopi ratios with values well above those measured at forward rapidities.
At large \pt ($> 3$ GeV/$c$) THERMINATOR fails to describe the data.  
The mismatch is clearly seen in the
 $\eta = 3.05$ plot of Fig. \ref{fig5}, where the data have the best \pt coverage. This mismatch is 
attributed to the fact that the model does not include
the production of jets.  
The AMPT model can qualitatively describe the pseudorapidity trends, but fails to 
quantitatively describe the data, namely, the model
under predicts \ptopip and over predicts \aptopi ratios. 
We note that the AMPT describes the \ptopip ratios
reasonably well for semi-peripheral reactions (not shown).

\section{Summary}

We present the $p_{T}$-dependence of the 
\ptopi ratios measured in Au+Au and p+p collisions
at energies of 62.4 and 200~GeV as a function of pseudorapidity and collision
centrality. The data provide
the opportunity to study baryon-to-meson production over a wide range
of the baryo-chemical potential $\mu_{B}$. 
For Au+Au collisions at \rootsnn{200} the \ptopip and \aptopi
ratios show a noticeable 
dependence
on centrality at intermediate \pt
with a rising trend from p+p to central Au+Au collisions. We 
find that \ptopip ratios are remarkably similar for central Au+Au at $\eta \approx 2.2$ at
\rootsnn{200}, and central Au+Au at  $\eta \approx 0$ at \rootsnn{62.4},  where the bulk medium is
characterized by the same value of $\bar{p}/p$. This observation,
together with the observed centrality dependence suggests
that particle production for intermediate \pt values is governed by the size
and the chemical properties of the created medium for the systems and pseudorapidity range studied.
It is also shown that the statistical model 
describes well the mid-rapidity \ptopip and \aptopi ratios for central
Au+Au at \rootsnn{200}, 
and it also tracks well the data trends at high rapidity.
Finally,  the Au+Au and p+p measurements at
\rootsnn{62.4} show that the \ptopip ratios for p+p and for all analyzed Au+Au
centralities cross simultaneously at the same $\eta$
value ($\approx$ 3.2) and are consistent with each other in the covered \pt range, e.g., 
from 0.3~GeV/$c$ up to 1.8~GeV/$c$. 
 
\section*{Acknowledgments}

   This work was supported by the Division of Nuclear Physics of the
Office of Science of the U.S. Department of Energy under contracts
DE-AC02-98-CH10886, DE-FG03-93-ER40773, DE-FG03-96-ER40981, and
DE-FG02-99-ER41121, the Danish Natural Science Research Council,
the Research Council of Norway, the Polish Ministry of Science and Higher
Education (Contract no. 1248/B/H03/2009/36), and the Romanian
Ministry of Education and Research (5003/1999, 6077/2000), 
and a sponsored research grant from Renaissance Technologies Corp.
We thank the staff of the Collider-Accelerator Division and the RHIC
computing facility at BNL for their support to the experiment.

\end{document}